# Snapping Mechanical Metamaterials under Tension

Ahmad Rafsanjani[1], Abdolhamid Akbarzadeh[1,2] and Damiano Pasini[1]*

[1] Mechanical Engineering Department, McGill University,
817 Sherbrooke Street West, Montreal, QC H3AOC3, Canada
[2] Bioresource Engineering Department, McGill University,
21111 Lakeshore Road, Ste-Anne de-Bellevue, Island of Montreal QC H9X 3V9, Canada

ABSTRACT We present a monolithic mechanical metamaterial comprising a periodic arrangement of snapping units with tunable tensile behavior. Under tension, the metamaterial undergoes a large extension caused by sequential snap-through instabilities, and exhibits a pattern switch from an undeformed wavy-shape to a diamond configuration. By means of experiments performed on 3D printed prototypes, numerical simulations and theoretical modeling, we demonstrate how the snapping architecture can be tuned to generate a range of nonlinear mechanical responses including monotonic, S-shaped, plateau and non-monotonic snap-through behavior. This work contributes to the development of design strategies that allow programming nonlinear mechanical responses in solids.

Mechanical metamaterials are man-made materials, usually fashioned from repeating unit cells which are engineered to achieve extreme mechanical properties, often beyond those found in most natural materials.[1] They gain their unusual, sometimes extraordinary, mechanical properties from their underlying architecture, rather than the composition of their constituents. Metamaterials exhibit interesting mechanical properties, such as negative Poisson's ratio,[2,3] negative incremental stiffness,[4] negative compressibility[5] and unusual dynamic behavior for wave propagation.[6] As Ron Resch (artist and applied geometrist) points out in his statement "the environment responds by collapsing quite often",[7] instabilities can be exploited to design advanced materials with innovative properties.[8] Recently, harnessing elastic instabilities played a central role in the rational design of novel 2D[9-14] and 3D[15-17] mechanical metamaterials with either significantly enhanced mechanical properties or equipped with new functionalities, e.g. programmable shape transformations.[14] In most of the examples mentioned above, elastic instabilities are exploited to trigger a pattern switch by a broken rotational symmetry, mostly governed by Euler buckling. In these works instabilities are induced by an applied compressive load.[16] This observation naturally leads to the question of whether one can either benefit from other mechanical instability mechanisms for metamaterial design or extend current concepts to other loading conditions.

In this work, we exploit mechanical instabilities triggered by *snap-through buckling* to create a metamaterial which experiences a pseudo pattern switch in *tension* and exhibits a programmable mechanical response. Our design is inspired by a monolithic bistable mechanism,[18] i.e. two curved parallel beams that are centrally-clamped as schematized in **Figure 1**a. A normal force applied in the middle of the double-beam mechanism can prompt it to snap through to its second stable state (Figure 1a, dashed lines). We release clamped conditions at both ends to create a repeatable unit cell (Figure 1b), composed of two centrally connected cosine-shaped slender segments, which can be tessellated in plane to form a periodic arrangement. When pulled along its axis of symmetry (*y*-axis), at a critical tensile strain $\varepsilon_{cr}$, the lower segment snaps through and the structure exhibits a pattern switch from a wavy-shaped structure to a diamond-like configuration (Figure 1c-e). Depending on the amplitude of the curved segments, this transition can be either smooth or discontinuous. The amplitude of the cosine-shaped curved segment can thus provide a means to tune the mechanical response of the system. We showcase the above-mentioned concepts with experiments on the fabricated prototypes and examine the robustness of our findings by performing finite element simulations and a theoretical analysis.

*Corresponding Author (damiano.pasini@mcgill.ca)

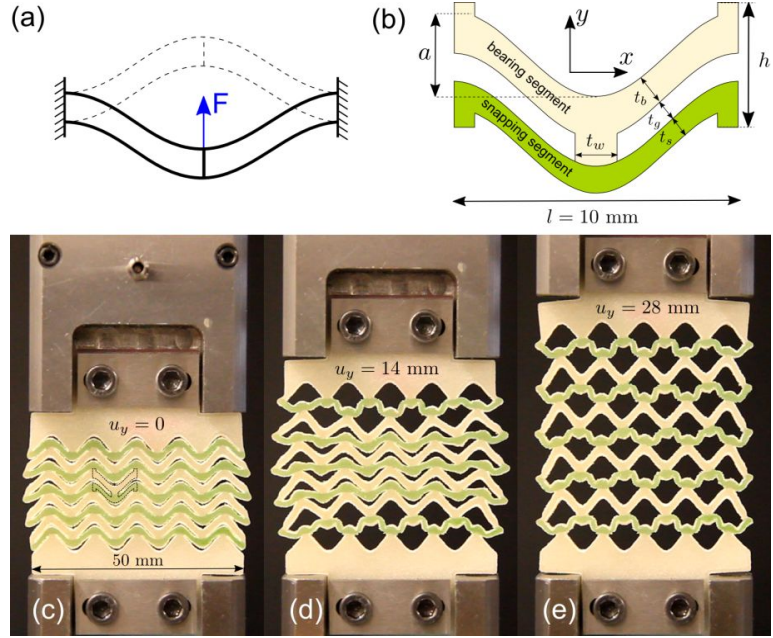

**Figure 1.** (a) Bistable mechanism of double curved beams which can snap between two stable configurations, under a vertical force applied in the middle (adopted from Qiu *et al.*[18]) (b) Unit cell geometry of the designed metamaterial composed of load bearing and snapping segments. The snapping segments are painted by a color marker after 3D printing for clarity. Under tension, the snapping segments snap through and the unit cell switches from an undeformed wavy-shape to a diamond configuration. (c-e) Snapshots of the 3D printed snapping mechanical metamaterial comprised of 5×5 unit cells ($l = 10\ mm$, $t_b = t_w = 1.5\ mm$ and $t_s = t_g = 1\ mm$) in response to tensile loading: (c) undeformed state ($u_y = 0$), (d) during snapping ($u_y = 14\ mm$) and (e) at full extension ($u_y = 28\ mm$).

We perform experiments on 3D printed specimens (Shapeways, NY, USA) fabricated using selective laser sintering technology (EOS e-Manufacturing Solutions, Germany) from a Nylon-based rubber-like material (Young's modulus of $E \simeq 78\ MPa$ and Poisson's ratio $\nu \simeq 0.4$). The specimens comprised of an array of 5×5 unit cells. Three different amplitude sizes ($a/l = 0.2, 0.3, 0.4$) are considered, whereas all the other geometrical parameters are kept constant ($l = 10\ mm$, $t_b = t_w = 1\ mm$ and $t_s = t_g = 1\ mm$). The out-of-plane thickness of all samples is $b = 3\ mm$. The structures are pulled in a uniaxial testing machine (Bose ElctroForce 3510) at a constant rate of $\dot{u}_y = 1\ mm/s$. The pulling forces and displacements are acquired, while a high-resolution digital camera facing the specimen records a video.

**Figure 2**a-c shows the nominal stress-strain curves (calculated as $\sigma_y = F_y/(n_x\ b\ l)$ and $\varepsilon_y = u_y/(n_y\ h)$, where $F_y$ is the reaction force, $u_y$ the displacement along y-axis and $n_x = n_y = 5$) of the snapping mechanical metamaterials for given amplitude parameters $a/l$ in response to a tensile load. The stress-strain curves exhibit three regions. (i) A small strain regime, where the specimens respond linearly up to a critical strain $\varepsilon_{cr} = 0.2\sim0.25$. In this phase, the response is governed by bending-dominated deformations in the snapping segments. (ii) Snapping strain regime, where a further stretching triggers elastic instabilities. The snapping segments snap through row-by-row, with relatively small changes in the tensile load, until the whole specimen is fully stretched. Here, for the smaller value of the amplitude parameter $a/l = 0.2$, the stress-strain curve reaches a plateau, whereas for larger amplitudes, i.e. $a/l = 0.3, 0.4$, the response is non-monotonic. During snapping, the slope of the stress-strain curves becomes negative, a phenomenon described by a negative incremental stiffness.[4] (iii) Stiffening regime, where the cell walls orient along the loading direction, and the slope of the stress-strain curve rises again due to the transition into a fully stretching-dominated deformation state.

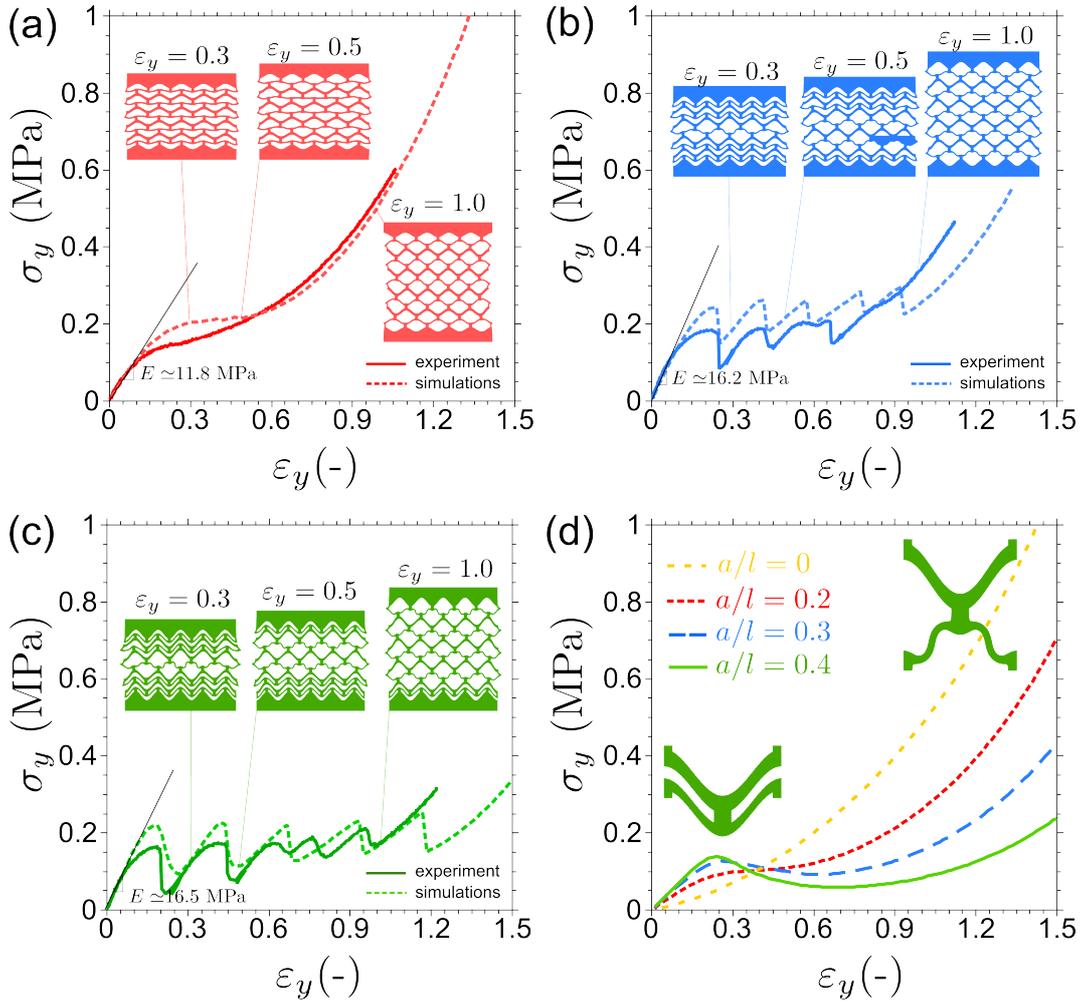

**Figure 2.** Nominal stress-strain responses from experiments and FEA simulations for the snapping mechanical metamaterial composed of 5×5 unit cells for the following amplitude parameters (a) $a/l = 0.2$, (b) $a/l = 0.3$ and (c) $a/l = 0.4$. For small amplitudes of $a/l$, the response is smooth without snap-through instabilities, whereas for larger values of $a/l$, the metamaterial snaps sequentially and exhibits a discontinuous response. (d) Mechanical response of a single unit cell for selected $a/l$ parameters.

The response of the snapping metamaterial under uniaxial tensile load is further investigated by performing a finite element analysis (FEA) of a full scale and a unit cell model using nonlinear FEA package ABAQUS, where all parameters match those of the experiments. The models are meshed using six-node triangular, quadratic plane stress elements (type CPS6) and mesh sensitivity analysis is performed to ensure accuracy. A neo-Hookean hyperelastic material is assigned and the material parameters are adopted from the experimentally measured stress-strain curves of the constitutive material. A good quantitative agreement is achieved between experiments and simulations. As reflected in Figure 2a-c, all three regions observed in the experimental stress-strain curves are reproduced in the simulations, which further establishes the reliability of our computational approach. Representative examples of the simulated deformed states of each specimen are illustrated in the inset. The amplitude parameter $a/l$ is found to affect significantly both the stress-strain relationship and the effective elastic modulus, although all the specimens are assumed to possess approximately identical porosity. Figure 2d illustrates the mechanical response of a single unit cell showing similar behavior to those observed in the experiments and full scale simulations. Here, the repeated snaps are not present, since only a single unit cell is considered. The number of rows of the unit cells along the $y$-direction governs the nonlinear mechanical response, i.e. the number of snaps, as demonstrated in the supplementary information.[19] For $a/l = 0$, the response is monotonic and the snapping behavior vanishes.

**Figure 3** shows the deformation patterns of each specimen during the three response regimes allowing a comparison with their fully stretched configurations obtained via FEA. As observed in Figure 2b-c, for large amplitudes $a/l$, the instabilities localize and the structure exhibits a discontinuous deformation, whereas for smaller amplitudes (Figure 2a), a smooth transition appears. A supplementary video is provided which demonstrates the deformation of the snapping metamaterials in the experiments and FEA simulations.

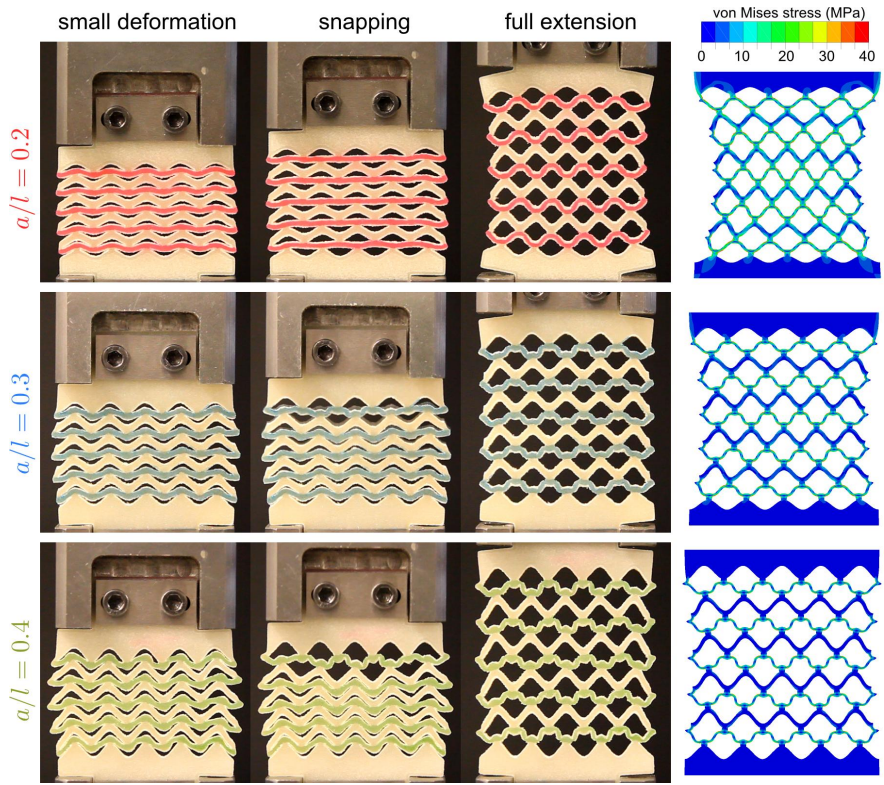

**Figure 3.** Snapshots of the snapping mechanical metamaterials at small deformation, onset of snapping and full extension phases for three amplitude sizes ($a/l = 0.2, 0.3, 0.4$) of the curved segments. The last column shows the FEA predictions of the deformed metamaterial at full extension and the von Mises stress distributions. All samples are 3D printed from a single rubber-like material. The snapping segments are painted by a color marker for clarity.

We systematically explore the response of unit cells under uniaxial tension to gain further insight into the mechanical responses of the metamaterial under investigation. The unit cells are stretched along the *y*-axis, whereas the deformation along the *x*-axis is restrained. To reduce the number of the parameters involved, the thickness parameters defined in Figure 1b are assumed to be identical, i.e. $t_b = t_s = t_w = t_g = 0.1\,l$, which implies $h = 4t$. The resulting stress-strain curves from 4800 simulations in the ranges $0 \leq a/l \leq 0.48$ and $0.04 \leq t/l \leq 0.16$ are classified and a phase diagram is derived in the parameter space $(a/l, t/l)$ as shown in **Figure 4**a. The stress-strain curves fall into three categories with (i) strictly monotonic, (ii) S-shaped and (iii) non-monotonic snapping responses, with representative examples for each category illustrated in the inset. Structures with small amplitudes or thick walls show a strictly monotonic stress-strain curve. The S-shaped and the snapping responses are characterized by a non-monotonic variation of the incremental stiffness. For the S-shaped responses, the incremental stiffness is positive throughout the entire range of extension, as opposed to the snapping responses where the incremental stiffness becomes negative when snapping occurs. The boundary (dashed line), separating the S-shaped curves from the snapping responses, corresponds to configurations with a plateau in their stress-strain curves, i.e. a zero incremental stiffness appears over a finite range of extension before a positive stiffness is again observed.

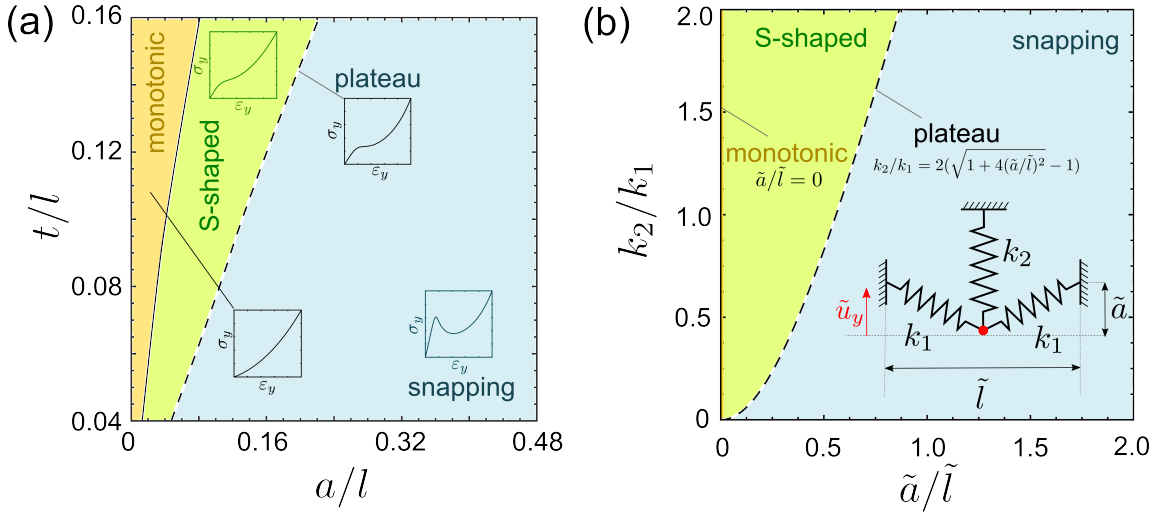

**Figure 4.** Phase diagrams for mechanical responses of (a) unit cells under uniaxial extension in the parameter space $(a/l, t/l)$, (b) for a single degree of freedom soft spring mechanism in the parameter space $(k_2/k_1, \tilde{a}/\tilde{l})$ showing monotonic, S-shaped and snapping responses and the plateau region.

To better understand the nature of the mechanical behavior observed in our experiments and simulations, a soft spring model is developed. The mechanism stores the elastic strain energy via three connected elastic springs with constants $k_1$ and $k_2$ (Figure 4b). The springs are initially unstressed, and their free ends are connected to the walls by joints, which allow rotation and restrain translation. The inclined springs ($k_1$) stand for the stiffness of the snapping segments, whereas the vertical spring ($k_2$) represents the interaction of the snapping segments with the rest of the structure. The geometry of this mechanism is characterized by the parameters $\tilde{l}$ and $\tilde{a}$, qualitatively equivalent to the parameters $l$ and $a$ of the metamaterial shown in Figure 1b. The state of this mechanism is governed by the vertical position of the middle joint, where an applied vertical force $P$ results in a vertical displacement $\tilde{u}_y$. The total potential energy of the conservative system is $W = U + \Pi$, which consists of the elastic strain energy $U$, stored by the springs, and the external-force potential $\Pi$. The equilibrium state of the system can be achieved from the stationary condition ($\delta W = 0$) of the total potential energy (please see supporting information). In this case, the force $P$ is obtained by:

$$P = 2k_1 \left(1 - \frac{\sqrt{\tilde{a}^2 + \frac{\tilde{l}^2}{4}}}{\sqrt{(\tilde{a} - \tilde{u}_y)^2 + \frac{\tilde{l}^2}{4}}}\right)(\tilde{u}_y - \tilde{a}) + k_2 \tilde{u}_y \tag{1}$$

Figure 4b shows the phase diagram of the soft spring system in the parameter space $(k_2/k_1, \tilde{a}/\tilde{l})$. For $k_2/k_1 = 0$, one could expect a pure snapping behavior in the inclined springs, whereas S-shaped and plateau responses appear as $k_2/k_1$ increases. The plateau surface can be calculated as $k_2/k_1 = 2(\sqrt{1 + 4\tilde{a}^2/\tilde{l}^2} - 1)$. Here, the monotonic behavior occurs for $\tilde{a}/\tilde{l} = 0$, which corresponds to the response of the unit cells with very thin walls and small amplitudes of cosine-shape. The resulting phase diagram for the soft spring system is qualitatively similar to the one obtained for unit cells. All types of observed behavior including monotonic, S-shaped, plateau and snapping responses are captured well with this single degree of freedom, spring model.

We have studied a novel class of mechanical metamaterials whose tensile stress-strain curve can be tuned by harnessing snap-through instabilities. If pulled along the *y*-axis, it snaps sequentially from an initially wavy pattern through several metastable states till a full extension is reached. This performance makes it potentially suitable for the design of morphing, adaptive, and deployable structures. The energy dissipated of snap-through phenomenon can be used for vibration isolation and damping applications.[20] This metamaterial can also serve as an adjustable tensile device for wave guide applications in photonic/phononic crystals.[21] The soft spring mechanism suggests that the interaction of the snapping segments with the rest of the structure can lead to mechanical responses with specific characteristics including incremental positive, zero or negative stiffness. The experimental results and FEA predictions exhibit quantitative agreement throughout the extension range, in both the shape of the stress-strain curves and the nature of the deformation patterns. Snap-through instabilities have been employed for material design in bistable periodic structures under compressive load,[22, 23] and for tuning surface topography.[24] More, recently, Shan *et al.*[25] designed an architected material that exhibits multi-stable energy-absorbing behavior due to snap-through instability in thin beam like elements when subjected to compression. However, to the best authors' knowledge, this is the first time that snap-through buckling is embedded in mechanical metamaterials under tension that allows remarkably large strain up to 150%. In particular, all segments in our designed metamaterial have approximately uniform thickness; a key feature that promotes more evenly distributed stresses and therefore prevents unpredicted failure. With several others, this work contributes to pave the way towards the study of nonlinear mechanical response of mechanical metamaterials with new functionalities.

**Acknowledgements** A.R. acknowledges the financial support provided by Swiss National Science Foundation (SNSF) under grant no. 152255. A.A. acknowledges the financial support and postdoctoral fellowship by Natural Sciences and Engineering Research Council (NSERC) of Canada.

# Supporting Information

Effect of the number of unit cells on the snapping behavior

A parametric study is performed to investigate the role of the number of unit cells on the response of the snapping metamaterial. We consider arrays of 1×1 to 5×5 unit cells with periodic boundary conditions under uniaxial extension and present the stress-strain responses in Figure S1. The material properties and the geometrical parameters are isolated to the parameters used in Figure 2. For small amplitude of the parameter $a/l = 0.2$, no difference can be observed in the stress-strain curve. However, for larger amplitudes, i.e. $a/l = 0.3, 0.4$, while the unit cell responses are identical in the small strain and large strain (stiffening) regimes, they differ during the snapping phase. In this case, the number of rows governs the occurrence of the sequential snapping, which shows the role of the representative volume element size in the mechanical response of a periodic material.[19]

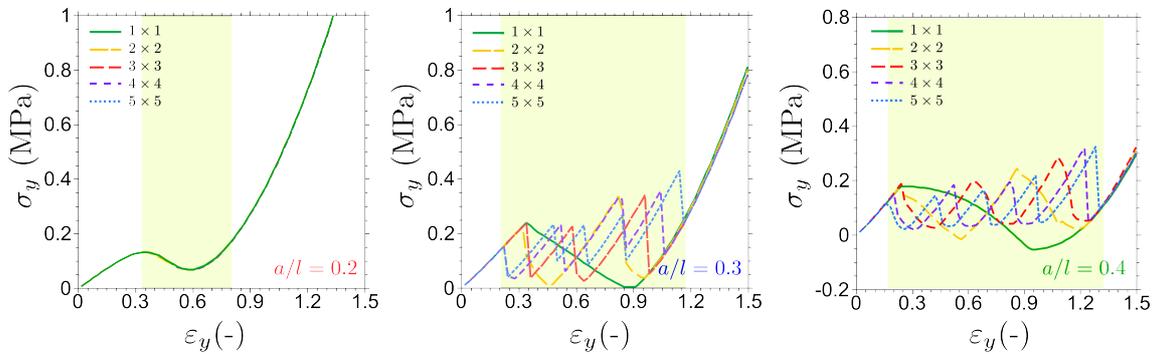

**Figure S1**. Nominal stress-strain curves of arrays of 1×1 to 5×5 unit cells under uniaxial extension for amplitude parameters $a/l = 0.2, 0.3, 0.4$. The highlighted regions correspond to the snapping regime.

Incremental negative stiffness vs. negative stiffness

By definition, a *negative stiffness* appears when the slope of the stress-strain curve is negative at zero force, whereas a negative slope at a non-zero load of a non-monotonic stress-strain curve represents a *negative incremental stiffness* (Moore *et al.*, *Phil. Mag. Lett.* 2006, 86, 651).

As shown in the phase diagram of Fig. 4a, the $a/l$ ratio has a significant effect on the mechanical response of the material. The stress-strain curves in the snapping phase, initially exhibit a *negative incremental stiffness* and further increasing the $a/l$ ratio results in *negative stiffness* behavior. We consider a single unit cell to demonstrate these phenomena. As shown in Fig. S2, for $a/l = 0.1$ the stress-strain curve exhibits positive stiffness everywhere while for $a/l = 0.2$ and $a/l = 0.4$, negative incremental stiffness and negative stiffness behaviors are detected, respectively. Negative stiffness represents a local minimum in the energy landscape and the resulting metamaterial is therefore multistable.

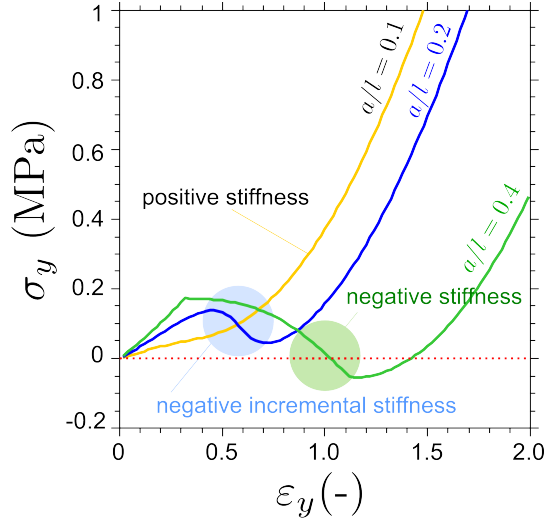

**Figure S2.** Nominal stress-strain curves for a unit cell ($l = 10\ mm$ and $t_b = t_w = t_s = t_g = 1\ mm$) under uniaxial extension exhibiting positive stiffness ($a/l = 0.1$), negative incremental stiffness (highlighted blue region for $a/l = 0.2$) and negative stiffness (highlighted green region for $a/l = 0.4$).

It is noteworthy that the unit cells are laterally restrained which anticipates snapping response. Finite size samples are not laterally confined and we need therefore relatively a larger $a/l$ ratio to achieve negative incremental stiffness and negative stiffness behaviors.

Soft spring model

As shown in the inset of Figure 4b, a soft spring model is presented to better elucidate the observed phenomena in experiments and simulations. The mechanism stores the elastic strain energy via three connected linear elastic springs. The two inclined springs ($k_1$) represents the snapping segments, whereas the vertical spring ($k_2$) stands for the interaction of the snapping segments with the rest of the metamaterial. The total potential energy of the conservative system consists of the elastic strain energy $U$, stored by the springs, and the external force potential $\Pi$, i.e. $W = U + \Pi$ written as:

$$W = k_1 \left( \sqrt{(\tilde{a} - \tilde{u}_y)^2 + \frac{\tilde{l}^2}{4}} - \sqrt{\tilde{a}^2 + \frac{\tilde{l}^2}{4}} \right)^2 + \frac{1}{2} k_2 \tilde{u}_y^2 - P \tilde{u}_y \tag{S1}$$

The equilibrium path of the system can be reached via the stationary condition of the total potential energy, i.e. $\delta W = 0$. In this case, the force ($P$)–displacement ($u_y$) relation is obtained by:

$$P = 2k_1 \left( 1 - \frac{\sqrt{\tilde{a}^2 + \frac{\tilde{l}^2}{4}}}{\sqrt{(\tilde{a} - \tilde{u}_y)^2 + \frac{\tilde{l}^2}{4}}} \right) (\tilde{u}_y - \tilde{a}) + k_2 \tilde{u}_y \tag{S2}$$

This equation provides the response of the soft spring model in terms of the stiffness constants and geometrical features. Several force-displacement (stress-strain) scenarios, presented in the phase diagram of the Figure 4, can be distinguished via the first and second derivatives of the force with respect to the displacement. In particular, the plateau surface, in Figure 4b, corresponds to the responses that exhibit a stationary point of inflection at $\tilde{u}_y = \tilde{a}$ and for which $P' = P'' = 0$, leading to:

$$\frac{k_2}{k_1} = 2\left(\sqrt{1 + 4\left(\frac{\tilde{a}}{\tilde{l}}\right)^2} - 1\right) \tag{S3}$$

To better refine the soft spring model, a multi-degree of freedom spring system could be used.